\input{aipcheck}

\documentclass[
    ,final            
  ]
  {aipproc}

\layoutstyle{6x9}

\begin{document}

\title{How Universal Is the Coupling in the Sigma Model?}


\author{Denis Parganlija}{
  address={Institut für Theoretische Physik, Johann Wolfgang Goethe-Universit\"at, Max von Laue-Str. 1, D-60438 Frankfurt am Main, Germany}
}

\author{Francesco Giacosa}
{
 address={Institut für Theoretische Physik, Johann Wolfgang Goethe-Universit\"at, Max von Laue-Str. 1, D-60438 Frankfurt am Main, Germany}
}

\author{Dirk H. Rischke}{
  address={Institut für Theoretische Physik and Frankfurt Institut for Advanced Studies, Johann Wolfgang Goethe-Universit\"at, Max von Laue-Str. 1, D-60438 Frankfurt am Main, Germany}
}

\begin{abstract}
We calculate pion-pion scattering lengths and sigma, rho and $a_1$ decay widths from a gauged linear sigma model with two flavours and its globally invariant generalisation.
\end{abstract}

\maketitle


\section{INTRODUCTION}

Quantum Chromodynamics (QCD) at low energies can
be successfully described by means of effective approaches which display
the same global symmetries of QCD, most notably the chiral $SU(N_{f})_{r}\times SU(N_{f})_{l}$ symmetry, where $N_{f}$ is the number of
flavors. Effective models are expressed in terms of hadronic degrees of
freedom and not in terms of quarks and gluons. Spontaneous breaking of
chiral symmetry implies that the pseudoscalar mesons (i.e., the pions) emerge
as almost massless Goldstone bosons and the scalar states which are the
corresponding chiral partners acquire a large mass.
We distinguish between the nonlinear and linear realizations of chiral
symmetry: while in the nonlinear case the scalar excitations are integrated out
and only pseudoscalar mesons, interacting via derivative couplings, are
left, in the linear case both scalar and pseudoscalar degrees of freedom are present.

In this paper we work with the latter by using a generalized linear sigma
model in which, besides scalar and pseudoscalar mesons, also vector and
axial-vector degrees of freedom are included. We first revisit the
hypothesis of local chiral symmetry as described in Refs.\ \cite{GG,KR} and
show that a successfull description of pion-pion scattering and some important decay widths cannot be achieved simultaneously in this framework. As outlined in Ref.\ \cite{UBW}, allowing for global chiral symmetry
represents a viable extension: a first study in this direction is performed
here in the case of $N_{f}=2$.

The scalar fields entering the model are interpreted as quark-antiquark
states in agreement with large-$N_{c}$ counting rules. We then have two
possible scenarios: (a) the resonances below 1 GeV $f_{0}(980)$, $a_{0}(980)$, $k(800)$ and $f_{0}(600)$ represent the quarkonia nonet. Thus, the states $f_{0}(600)$ and $a_{0}(980)$ are identified with the $\sigma $ and the $a_{0}$ fields of our $N_{f}=2$ model. (b) The quarkonia are heavier
than 1 GeV: the resonances $f_{0}(1370)$, $f_{0}(1500)$, $f_{0}(1710)$, $a_{0}(1450)$, $K_{0}(1430)$ describe a full nonet, in which the isoscalar
states mix with the glueball \cite{refs1}. In the case $N_{f}=2$ the
resonances $f_{0}(1370)$ and $a_{0}(1450)$ correspond to the $\sigma $ and
the $a_{0}$ fields. The scalars below 1 GeV, whose spectroscopic wave
functions possibly contain a dominant tetraquark or mesonic molecular
contribution in this scenario \cite{refs2, Schechter}, may be introduced in the model as
extra scalar fields. In this work we briefly outline how we intend to
explore both scenarios (a) and (b) in the future.

\section{THE MODEL AND ITS IMPLICATIONS}
\emph{The gauged linear sigma model:} The Lagrangian of the gauged linear
sigma model with $U(2)_{R}\times U(2)_{L}$ symmetry reads \cite{GG}: 
\begin{eqnarray}
\lefteqn{ {\cal L}=\,\mbox{Tr}[(D^{\mu } \Phi )^{\dagger }(D^{\mu } \Phi
)]- m_{0}^{2}\,\mbox{Tr}(\Phi ^{\dagger }\Phi )-\lambda _{1}[\,\mbox{Tr}(\Phi
^{\dagger }\Phi )]^{2} - \lambda _{2}\,\mbox{Tr}(\Phi ^{\dagger }\Phi )^{2}} 
\nonumber \\
&&-\,\frac{1}{4}\,\mbox{Tr}[(F_{l}^{\mu \nu })^{2}+(F_{r}^{\mu \nu })^{2}]+%
\frac{m_{1}^{2}}{2}\,\mbox{Tr}[(A_{l}^{\mu })^{2}+(A_{r}^{\mu })^{2}]+%
\mbox{Tr}[H(\Phi +\Phi ^{\dagger })]  \nonumber \\
&&+\,c\,(\mbox{det}\,\Phi +\mbox{det}\,\Phi ^{\dagger })\,,\qquad \qquad
\qquad \qquad \quad   \label{Lagrangian}
\end{eqnarray}
with $\Phi =(\sigma +i\eta )\,t^{0}+(\vec{a}_{0}+i\vec{\pi})\cdot \vec{t}$
(scalar and pseudoscalar mesons), $A_{l,r}^{\mu }=(\omega ^{\mu }\pm
f_{1}^{\mu })\,t^{0}+(\vec{\rho}^{\mu }\pm \vec{a}_{1}^{\mu })\cdot \vec{t}$
(vector and axialvector mesons), where $t^{0}$, $\vec{t}$ are the generators
of $U(2)$; $D^{\mu }\Phi =\partial ^{\mu }\Phi +ig(\Phi A_{l}^{\mu
}-A_{r}^{\mu }\Phi )$ and $F_{l,r}^{\mu \nu }=\partial ^{\mu }A_{l,r}^{\nu
}-\partial ^{\nu }A_{l,r}^{\mu }-ig\,[A_{l,r}^{\mu },A_{l,r}^{\nu }]$.
Explicit symmetry breaking is described by the term Tr$[H(\Phi +\Phi
^{\dagger })]\equiv h\sigma (h=const.)$ and the chiral anomaly by the term $c\,(\mbox{det}\,\Phi +\mbox{det}\,\Phi ^{\dagger })$ \cite{Hooft}. Note that
the local symmetry in the Lagrangian (\ref{Lagrangian}) is explicitly broken
to a global symmetry by non-vanishing vector meson masses. The respective
term gives rise to the celebrated current-field proportionality \cite{GG}.

The explicit form of the Lagrangian after spontaneous symmetry breaking can
be found, e.g., in Ref.\ \cite{RS}. Note that all the parameters in the
Lagrangian are fixed by the tree-level meson masses and the pion decay
constant $f_\pi$.\\

\begin{figure}[h]
\includegraphics[scale = 0.14]{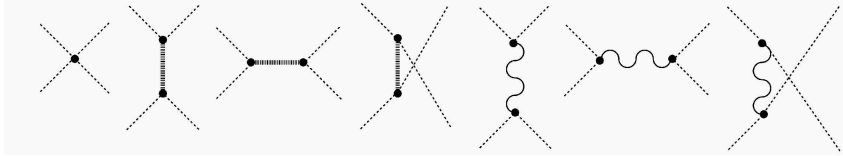}
\caption{Tree-level pion-pion scattering (dashed line: pions; solid line:
sigma; wavy line: rho mesons)}
\label{Figure}
\end{figure}

\emph{Scattering lengths:} In order to calculate the scattering lengths, at
tree level one has to compute the amplitude corresponding to the diagrams
shown in Fig.\ \ref{Figure}. 

Partial wave decomposition \cite{Leutwyler} then leads to the following expressions for the scattering lengths in units of the pion mass:
\begin{eqnarray}
& a_{0}^{0}=\frac{m_{\pi}^{2}}{32\pi f_{\pi}^{2}} \, \left [7 + \frac{2}{Z^2}%
\frac{m_{\pi}^{2}}{m_{\sigma}^{2} }\, \left (1-\frac{2}{Z^{2}} \right )^{2}
+ \frac{3}{Z^{2}}\frac{m_{\pi}^{2} }{m_{\sigma}^{2} - 4m_{\pi}^{2}} \, \left
(1+\frac{2}{Z^{2}} \right )^{2} \right ]\, ,&  \label{a00} \\
& a_{0}^{2} = -\frac{m_{\pi}^{2}}{16 \pi f_{\pi}^{2}} \, \left [1 - \frac{1}{%
Z^{2}}\frac{m_{\pi}^{2}}{m_{\sigma}^{2}} \, \left (1 - \frac{2}{Z^{2}}
\right )^{2} \right ] &  \label{a02}
\end{eqnarray}
with $Z = \frac{m_{a_1}}{m_\rho}$ and $f_\pi = 92.4$ MeV. Note that the PDG
value is $m_{a_1}=1230$ MeV whereas the KSFR \cite{KSFR} rule suggests $m_{a_1}\stackrel{!}{=}\sqrt{2}m_{\rho} = 1097$ MeV.\\
Given the data on scattering lengths: $a_0^0=0.224\pm0.03$ and $a_0^2=-0.037\pm 0.024$ (NA48/2 Cusp) \cite{Wanke}; $a_0^0= 0.233 \pm 0.023$
and $a_0^2 = -0.047 \pm 0.015$ (NA48/2 Ke4) \cite{Wanke}, it follows from
Eqs.\ (\ref{a00}) and (\ref{a02}) that within the framework of the
Lagrangian (\ref{Lagrangian}) a light scalar meson with the mass $m_\sigma
\simeq (315 - 345)$ MeV is favoured, depending on the choice of $a_1$ mass
and thus the choice of the parameter $Z$. \\

\emph{Decay widths: }The results for the decay widths of the sigma
and the rho into two pions that follow from the Lagrangian (\ref{Lagrangian}) fail to reproduce experimental results \cite{PDG}. The calculation of the $\sigma \rightarrow \pi \pi $ decay width yields 
\begin{eqnarray*}
\Gamma _{\sigma \rightarrow \pi \pi }=\frac{3}{\,32\pi }\frac{m_{\sigma }^{3}}{\,Z^{6}f_{\pi }^{2}}\sqrt{1-\left( \frac{2m_{\pi }}{m_{\sigma }}\right)
^{2}}\left[ 1+\frac{\,m_{\pi }^{2}}{\,m_{\sigma }^{2}}(Z^{2}-2)\right] ^{2}.
\end{eqnarray*}
One obtains $\Gamma _{\sigma \rightarrow \pi \pi }<107$ MeV for $m_{\sigma
}<800$ MeV, clearly too small when compared to the PDG value which ranges
between 600 MeV and and 1.2 GeV.\\
The decay width of the $\rho$ into two pions is given by 
\begin{eqnarray*}
\Gamma _{\rho \rightarrow \pi \pi }= \frac{g^2 }{192 \pi} \, m_\rho \left [1 - \left ( \frac{2m_\pi}{m_\rho} \right )^2 \right ]^\frac{3}{2}  \left ( 1 + \frac{1}{Z^2} \right )^2; \;g=\frac{\sqrt{Z^2 -1}}{Z}\frac{m_\rho}{f_\pi} \approx 6.51.
\end{eqnarray*}
Then the value $\Gamma _{\rho \rightarrow \pi \pi } = 86.5$ MeV is obtained: almost a factor of two lower than the
experimental value $(149.4.\pm 1.0)$MeV. Additionally, it also follows from
Eq.\ (\ref{Lagrangian}) that $\Gamma _{a_{1}\rightarrow \rho \pi }\approx
300$ MeV, with the experimental value at $(250-600)$MeV.

\section{IMPROVEMENTS OF THE LAGRANGIAN}

A possible solution to the problem of the decay widths mentioned in the previous section was discussed in Refs.\ \cite{GG, KR, Meissner} where terms of higher dimension have been added to the Lagrangian. \\
However, as in Ref.\ \cite{UBW} we follow a different strategy: given that {\it i}) the local symmetry of the Lagrangian (\ref{Lagrangian}) is already broken to a global symmetry, and {\it ii}) there seems to be no reason why an effective field theory should have a local chiral symmetry if the same symmetry in the underlying theory (QCD) is a global one \cite{UBW}, one may promote the local symmetry in the Lagrangian (\ref{Lagrangian}) to a global one.\\
For a global chiral symmetry, up to scaling dimension four the following additional terms appear in the Lagrangian (\ref{Lagrangian}): Tr$(\Phi^\dagger \Phi)$Tr$(\vert A^\mu_l \vert^2 + \vert A^\mu_r \vert^2)$, Tr$(\Phi A_{\mu\, l} \Phi^\dagger A^\mu_r)$ and Tr$(\vert \Phi A^\mu_r \vert^2 + \vert A_l^\mu \Phi \vert^2)$ \cite{Boguta,KS,Pisarski}.
The consequences for the pion scattering lengths and decay widths arising
from the globally invariant Lagrangian are under investigation \cite{PGR}.
In this work we will restrict our discussion to noticing that global
invariance implies mathematically that the coupling constant in the
covariant derivative $D^{\mu }\Phi =\partial ^{\mu }\Phi +ig_{1}(\Phi
A_{l}^{\mu }-A_{r}^{\mu }\Phi )$ no longer needs to be the same as the one
appearing in the field strength tensor $F_{l,r}^{\mu \nu }=\partial ^{\mu
}A_{l,r}^{\nu }-\partial ^{\nu }A_{l,r}^{\mu }-ig_{2}\,[A_{l,r}^{\mu
},A_{l,r}^{\nu }]$; the ensuing division of a single coupling constant ($g$)
into two different ones means that the scalar-vector coupling ($g_{1}$) is
no longer the same as the vector-vector coupling ($g_{2}$) and at the same
time it provides us with a new parameter needed to adjust the $\rho
\rightarrow \pi \pi $ decay width to the experimentally very precisely
(within $\pm $1.0 MeV) observed value.\\

A recalculation of the $\rho \rightarrow \pi \pi $ width then yields
\begin{eqnarray*}
\Gamma_{\rho \rightarrow \pi\pi}= \frac{m_\rho}{48 \pi} \left [ 1 - \left (\frac{2m_\pi}{m_\rho} \right )^2 \right ]^\frac{3}{2}\,\left [ g_1 - \frac{g_2}{2} \left ( 1- \frac{1}{Z^2}  \right )  \right ]^2.
\end{eqnarray*}

Using the PDG value $\Gamma_{\rho \rightarrow \pi\pi}=149.4$ MeV one obtains\footnote{There also is a second solution $g_2 = 41.48$, which, however, leads to an unphysical (i.e., too large) value of the $a_1$ decay width into rho and pion.} $g_ 2 = 1.77$.\\

In the present model, the value of the $a_1 \rightarrow \rho \pi$ decay width depends on the value of the $a_1$ mass. Our results indicate that the $a_1$ resonance is broad (in accordance with the experiments); the exact value is $\Gamma_{a_1 \rightarrow \rho \pi} = 516$ MeV for $g_2 = 1.77$ and $m_{a_1} = 1097$ MeV (KSFR rule), and around 1.3 GeV for $m_{a_1} = 1230$ MeV. Improving the very large decay width value of $a_1$ at $m_{a_1}=1230$ MeV is currently under investigation \cite{PGR}. \\  

The value of the $\sigma \rightarrow \pi \pi$ decay width at threshold is not affected by the introduction of the new coupling constant $g_2$, hence it is still too small. To correct this, one or both of the following two steps may be taken: 
\begin{itemize}
\item Investigating all globally symmetric terms with no new states added \cite{PGR}; the $\sigma$ field is then interpreted as $f_0(600)$ and $a_0$ field as $a_0(980)$ and both are treated as quarkonia.
\item Assuming that the masses of scalar quarkonia are above 1 GeV, re-interpreting the $\sigma$ field as $f_0(1370)$ and $a_0$ field as $a_0(1450)$ and adding new terms corresponding to $f_0(600)$ and $a_0(980)$ (similar to the work in Refs. \cite{Schechter}). 
\end{itemize}
In this work we have taken the latter step and as a first solution to the $\sigma$ decay width problem we have added a scalar-pion interaction term
$
{\cal L}_{\tilde g,\,S} = \tilde g f_\pi S \, \left (\frac{\partial_\mu \vec \pi}{f_\pi} \right )^2
$,
where $S$ is now interpreted as $f_0(600)$, to the Lagrangian of Eq. (\ref{Lagrangian}). Therefore, two new parameters are introduced ($\tilde g$ and $m_S$) that may be adjusted to the two scattering lengths $a_0^0$ and $a_0^2$. Then the value of the pion-pion $f_0(600)$ decay width may be calculated.
A recalculation of the scattering lengths yields
\begin{eqnarray*}
a_{0}^{0 \, new} = a_{0}^{0}\,+\, \frac{\tilde g^{2}\,m_{\pi}^{4}}{\pi} \left (\frac{1}{m_{S}^{2}} - \frac{3}{2}\frac{1}{4m_{\pi}^{2}-m_{S}^{2}}  \right )\mbox{,} \;\; a_{0}^{2\, new} = a_{0}^{2}\,+\, \frac{\tilde g^{2}}{\pi} \frac{m_{\pi}^{4}}{m_{S}^{2}} \, ,
\end{eqnarray*}
with $a_0^0$ and $a_0^2$ from (\ref{a00}) and (\ref{a02}) respectively. 
Our best values for the $f_0(600)$ mass and decay width are at $m_S  = 608$ MeV and $\Gamma_{S \rightarrow \pi \pi}=466$ MeV (obtained for $a_0^0 =0.206$, $a_0^2 = -0.0295$ and $g_3 = 0.645$) which is within experimental values as quoted by the PDG \cite{PDG}.
This also means, however, that the quoted scattering lengths are within NA48/2 Cusp data and outside of NA48/2 Ke4 data \cite{Wanke}. This issue will be addressed in the future \cite{PGR}.

\section{CONCLUSIONS AND OUTLOOK}

A linear sigma model with vector and axial-vector mesons
has been utilized to study important processes of low-energy QCD. The necessity
and the consequences of abandoning local chiral symmetry and considering
globally symmetric terms have been discussed. In the future, a complete study of globally symmetric terms up to fourth order and their influence on
experimental results is required. In this way, we plan to address relevant issues concerning vacuum
phenomenology, such as the nature of scalar mesons and the inclusion of the
nucleon field together with its chiral partner \cite{Susanna}. Moreover, we plan to extend the work of Ref.\ \cite{RS} in order to consider chiral symmetry restoration at nonzero temperature.\\


\begin{theacknowledgments}
The authors thank S. Str\"uber for valuable discussions during the preparation of this work.
\end{theacknowledgments}




\bibliography{sample}


\end{document}